\documentclass[12pt, a4paper]{elsarticle}

\usepackage{amssymb}

\usepackage{pythonhighlight}
\journal{SoftwareX}

\usepackage{xspace}
\usepackage{graphicx}

\newcommand{\dasp}{Durham \ao simulation platform (DASP)\renewcommand{\dasp}{DASP\xspace}\renewcommand{\thedasp}{DASP\xspace}\renewcommand{\Thedasp}{DASP\xspace}\renewcommand{\daspcite}{DASP\xspace}\renewcommand{\daspcite}{DASP\xspace}\renewcommand{\thedaspcite}{DASP\xspace}\xspace}
\newcommand{\daspcite}{Durham \ao simulation platform \citep[DASP,]{basden5}\renewcommand{\dasp}{DASP\xspace}\renewcommand{\thedasp}{DASP\xspace}\renewcommand{\Thedasp}{DASP\xspace}\renewcommand{\daspcite}{DASP\xspace}\renewcommand{\thedaspcite}{DASP\xspace}\xspace}
\newcommand{\thedaspcite}{the Durham \ao simulation platform \citep[DASP,]{basden5}\renewcommand{\dasp}{DASP\xspace}\renewcommand{\thedasp}{DASP\xspace}\renewcommand{\Thedasp}{DASP\xspace}\renewcommand{\daspcite}{DASP\xspace}\renewcommand{\thedaspcite}{DASP\xspace}\xspace}

\newcommand{\thedasp}{the Durham \ao simulation platform (DASP)\renewcommand{\dasp}{DASP\xspace}\renewcommand{\thedasp}{DASP\xspace}\renewcommand{\Thedasp}{DASP\xspace}\renewcommand{\daspcite}{DASP\xspace}\renewcommand{\thedaspcite}{DASP\xspace}\xspace}
\newcommand{\Thedasp}{The Durham \ao simulation platform (DASP)\renewcommand{\dasp}{DASP\xspace}\renewcommand{\thedasp}{DASP\xspace}\renewcommand{\Thedasp}{DASP\xspace}\renewcommand{\daspcite}{DASP\xspace}\renewcommand{\thedaspcite}{DASP\xspace}\xspace}

\providecommand{\ao}{}
\renewcommand{\ao}{adaptive optics (AO)\renewcommand{\ao}{AO\xspace}\renewcommand{\Ao}{AO\xspace}\xspace}
\newcommand{\Ao}{Adaptive optics (AO)\renewcommand{\ao}{AO\xspace}\renewcommand{\Ao}{AO\xspace}\xspace}

\newcommand{\elt}{Extremely Large Telescope (ELT)\renewcommand{\elt}{ELT\xspace}\renewcommand{\elts}{ELTs\xspace}\renewcommand{\eelt}{European ELT (E-ELT)\renewcommand{\eelt}{E-ELT\xspace}\xspace}\xspace}

\newcommand{\elts}{Extremely Large Telescopes (ELTs)\renewcommand{\elt}{ELT\xspace}\renewcommand{\elts}{ELTs\xspace}\renewcommand{\eelt}{European ELT (E-ELT)\renewcommand{\eelt}{E-ELT\xspace}\xspace}\xspace}

\newcommand{\eelt}{European Extremely Large Telescope (E-ELT)\renewcommand{\eelt}{E-ELT\xspace}\renewcommand{\elt}{ELT\xspace}\renewcommand{\elts}{ELTs\xspace}\xspace}

\newcommand{\est}{European Solar Telescope (EST)\renewcommand{\est}{EST\xspace}\xspace}
\newcommand{\mpi}{the Message Passing Interface (MPI)\renewcommand{\mpi}{MPI\xspace}\xspace}

\newcommand{\dm}{deformable mirror (DM)\renewcommand{\dm}{DM\xspace}\renewcommand{\dms}{DMs\xspace}\renewcommand{\Dms}{DMs\xspace}\renewcommand{\Dm}{DM\xspace}\xspace}
\newcommand{\dms}{deformable mirrors (DMs)\renewcommand{\dm}{DM\xspace}\renewcommand{\dms}{DMs\xspace}\renewcommand{\Dms}{DMs\xspace}\renewcommand{\Dm}{DM\xspace}\xspace}
\newcommand{\Dms}{Deformable mirrors (DMs)\renewcommand{\dm}{DM\xspace}\renewcommand{\dms}{DMs\xspace}\renewcommand{\Dms}{DMs\xspace}\renewcommand{\Dm}{DM\xspace}\xspace}
\newcommand{\Dm}{Deformable mirror (DM)\renewcommand{\dm}{DM\xspace}\renewcommand{\dms}{DMs\xspace}\renewcommand{\Dms}{DMs\xspace}\renewcommand{\Dm}{DM\xspace}\xspace}

\newcommand{\psf}{point spread function (PSF)\renewcommand{\psf}{PSF\xspace}\renewcommand{\psfs}{PSFs\xspace}\renewcommand{\Psf}{PSF\xspace}\xspace}
\newcommand{\psfs}{point spread functions (PSFs)\renewcommand{\psf}{PSF\xspace}\renewcommand{\psfs}{PSFs\xspace}\renewcommand{\Psf}{PSF\xspace}\xspace}
\newcommand{\Psf}{Point spread function (PSF)\renewcommand{\psf}{PSF\xspace}\renewcommand{\psfs}{PSFs\xspace}\renewcommand{\Psf}{PSF\xspace}\xspace}
\newcommand{\lgs}{laser guide star (LGS)\renewcommand{\lgs}{LGS\xspace}\renewcommand{\Lgs}{LGS\xspace}\renewcommand{\lgss}{LGSs\xspace}\xspace}
\newcommand{\lgss}{laser guide stars (LGSs)\renewcommand{\lgs}{LGS\xspace}\renewcommand{\Lgs}{LGS\xspace}\renewcommand{\lgss}{LGSs\xspace}\xspace}
\newcommand{\Lgs}{Laser guide star (LGS)\renewcommand{\lgs}{LGS\xspace}\renewcommand{\Lgs}{LGS\xspace}\renewcommand{\lgss}{LGSs\xspace}\xspace}
\newcommand{\scao}{single conjugate \ao (SCAO)\renewcommand{\scao}{SCAO\xspace}\renewcommand{\Scao}{SCAO\xspace}\xspace}
\newcommand{\Scao}{Single conjugate \ao (SCAO)\renewcommand{\scao}{SCAO\xspace}\renewcommand{\Scao}{SCAO\xspace}\xspace}
\newcommand{\glao}{ground layer \ao (GLAO)\renewcommand{\glao}{GLAO\xspace}\xspace}
\newcommand{\Moao}{Multi-object \ao (MOAO)\renewcommand{\moao}{MOAO\xspace}\renewcommand{\Moao}{MOAO\xspace}\xspace}
\newcommand{\moao}{multi-object \ao (MOAO)\renewcommand{\moao}{MOAO\xspace}\renewcommand{\Moao}{MOAO\xspace}\xspace}
\newcommand{\mcao}{multi-conjugate adaptive optics (MCAO)\renewcommand{\mcao}{MCAO\xspace}\xspace}
\newcommand{\ltao}{laser tomographic \ao (LTAO)\renewcommand{\ltao}{LTAO\xspace}\xspace}

\begin{document}

\begin{frontmatter}



\title{The Durham Adaptive Optics Simulation Platform (DASP): Current status}


\author{A.\ Basden, N.\ A.\ Bharmal, D.\ Jenkins, T.\ J.\ Morris,
  J.\ Osborn, P.\ Jia, L.\ Staykov}

\address{Department of Physics, Durham University, South Road, Durham,
  DH1 3LE, UK}

\begin{abstract}

The Durham Adaptive Optics Simulation Platform (DASP) is a Monte-Carlo
modelling tool used for the simulation of astronomical and solar adaptive
optics systems.  In recent years, this tool has been used to predict
the expected performance of the forthcoming extremely large telescope
adaptive optics systems, and has seen the addition of several modules
with new features, including Fresnel optics propagation and extended
object wavefront sensing.  Here, we provide an overview of the features of
DASP and the situations in which it can be used.  Additionally, the
user tools for configuration and control are described.
\end{abstract}


\end{frontmatter}



The \dasp has been under
development since the early 1990s.  Its current framework was
established in 2006 to meet the challenges of modelling the
forthcoming extremely large telescopes, with primary mirror diameters
of over 20~m.  Since 2006, DASP has been regularly developed to improve
computational performance, increase simulation fidelity, and expand the number of features that can be modelled.  It uses a
modular design, allowing new developments and algorithms to be added
whilst maintaining compatibility.  DASP is developed primarily in Python
and C, and uses pthreads and MPI for parallelization enabling
modelling of the largest proposed telescopes on reasonable
timescales.  

\section{Motivation and significance}
\label{sect:motivation}

The Earth's atmosphere has a perturbing effect on incident starlight,
meaning that the effective spatial resolution of large telescopes
(typically anything larger than 20~cm diameter) is limited.  By using
\ao systems, the distorted wavefronts of incident light can be
measured and have a correction applied so that the effective
resolution is improved, thus enabling scientific observations to be
made.  Designing an \ao system to meet complex scientific requirements
is an involved process, and modelling of the system performance is
necessary.  Additionally, investigation of new algorithms, techniques
and concepts also requires verification by simulation.

\subsection{Scientific contribution of DASP}
\dasp was developed to meet the needs of \ao system designers, and
has previously been used to model the expected performance of several
of the forthcoming \elt instruments, including MOSAIC
\citep{basden12,basden15}, MAORY \citep{basden21} and HIRES.  Recent
developments have also introduced an extended object (wide
field-of-view) wavefront sensor module, enabling \dasp to be used for
the modelling of solar \ao systems (e.g.\
European Solar Telescope
\citep{2016SPIE.9908E..09M}).  Additionally, existing instruments have
also been modelled to demonstrate that proposed novel techniques can
improve the \ao system
performance \citep{basden14,basden23}.

\dasp enables \ao system designers to explore
the large parameter spaces associated with \ao system development,
allowing system design trade-offs to be made in an informed manner,
with typical parameters including \ao system order, number of guide
stars, and wavefront sensor pixel scale.  \dasp is used to design and
optimize new \ao systems, and to verify performance of existing
systems.  \dasp can be a crucial tool for understanding the \ao error
budget, allowing cost-effective decisions to be made about the design
optimizations that can be performed to allow a given design to meet
its required performance targets.

\dasp has seen adoption within the \ao community, and is now used for
forthcoming \elt instruments \citep{basden21,basden24}, for the 10.4~m
Gran Telescopio Canarias, the Kunlun Dark Universe Survey Telescope
\citep{2013RAA....13..875J}, the Chinese 2.16~m telescope
\citep{peng216} and for the proposed 12~m Chinese Large Optical
Telescope.

\subsection{Using DASP}
\dasp can be operated on any Linux computer, and also under the OS-X
operating system.  Once installed, the user will typically generate a
new simulation configuration using the daspbuilder tool, which allows
the user to select a number of configuration options covering most \ao
configurations, including \scao, \mcao, \glao, \ltao and \moao.  This
generates the necessary configuration files, which can then be edited
by the user.  Alternatively, for more complex simulations, a daspsetup
tool can be used to graphically design an \ao system.  This simulation
is then executed to model \ao performance.  A further description of
\dasp usage is given in Section \ref{sect:eg}

\subsection{Other AO simulations}
There are a number of other Monte Carlo \ao simulation tools freely
available to the community, including YAO \citep{2013aoel.confE..18R},
CAOS \citep{2005MNRAS.356.1263C}, SOAPY \citep{2016SPIE.9909E..7FR},
MAOS \citep{maos} and OOMAO \citep{2014SPIE.9148E..6CC}.  OCTOPUS
\citep{2016SPIE.9909E..75L} is another Monte Carlo \ao simulation
tool, which is available from the European Southern Observatory upon request.   However, none
of these offer the combined performance and extended object wavefront
sensing capabilities of \dasp.  In addition, a number of analytical
modelling tools also exist, including PAOLA \citep{Jolissaint:06} and
CIBOLA \citep{Ellerbroek:05}, though these tools are used for rapid
prototyping and do not offer high fidelity.






\section{Software description}
\label{sect:descr}

\begin{figure}
  \includegraphics[width=0.5\linewidth]{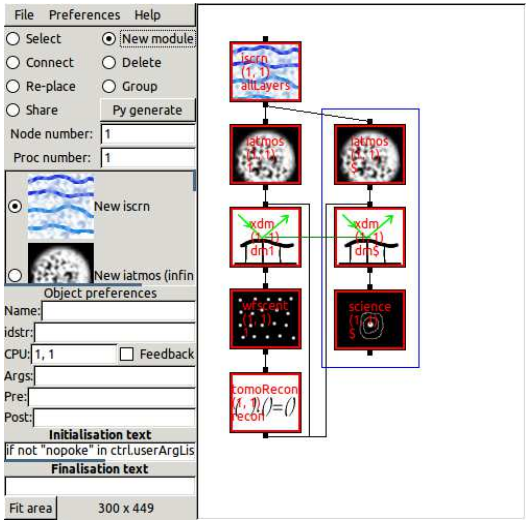}
  \caption{A screenshot of the daspsetup tool, showing the links
    between different science modules that create a simulation.  This
    tool can be used to configure an initial simulation, which is then
  executed from the command line.  Configuration of an AO simulation
  can be complicated when parallelization across multiple computing
  nodes is required, and the graphical tool simplifies this task.}
  \label{fig:daspsetup}
  \end{figure}

\dasp is comprised of a number of science modules, which model
discrete parts of an \ao system.  These include wavefront sensors
(including Shack-Hartmann and Pyramid sensors), deformable mirrors
(including zonal and modal), atmospheric models, wavefront
reconstruction modules, and astronomical object models.  This modular
design means that it is also possible for the user to add 
modules which can then be used during a simulation (for example,
another wavefront sensor type).  These science modules are linked
together to represent the flow of information through the \ao system,
as shown in Fig.~\ref{fig:daspsetup}.

\dasp also contains a number of utility modules, which are used by the
science modules, containing the necessary algorithms required by the
simulation.  User tools are also included, to analyse and display
results, to communicate with running simulations, and to configure
initial simulations.  \dasp offers good performance on laptops (for
simulation of smaller scale systems, typically up to 10~m class
diameter telescopes); and also HPC facilities for \elt
modelling. Inter-node communication is based on \mpi.

Once a simulation has been configured, it is run from the command
line.  After a set number of iterations (typically several thousand,
depending on \ao system type), the simulation then finishes, providing
the user with \ao performance metrics.  Each iteration represents one
simulation time-step, typically set by the integration time of
wavefront sensors, of order 1~ms.  It is usually necessary to model
tens of seconds of real time in order to ensure that results are
statistically valid.  The simulation will typically run between
10--1000 times slower than real-time, depending on scale and
complexity, though a simple system on a small (4~m) telescope can
operate faster than real-time.

Using computationally efficient code allows large \ao-related
parameter spaces to be explored within reasonable timescales.  For
example, a typical \elt-scale wide-field \ao simulation can be
completed within about 12 hours on a single server, where previously
this would have taken many days to complete.  This includes the
required system calibrations (e.g.\ interaction matrix generation and
reference slope computation), for a system with 6 laser guide stars
each with $80\times80$ sub-apertures and a control matrix of size
10~GB, with sufficient temporal averaging to smooth the atmospheric
turbulence effect (approximately 1~minute of \ao system time).

Top level modules are written in
Python, which allows rapid development of new algorithms and modules,
enabling users to tailor \dasp for their own use.  While a simulation
is running, it can be queried and modified using both command line and
graphical tools, useful for when initially configuring a simulation.
It also provides a way of rapidly determining stability conditions for
an \ao simulation before further extensive parameter space
exploration.

\subsection{Software Architecture}
\label{sect:arch}

The \dasp top-level directory structure separates the simulation into
logical components:
\begin{itemize}
\item {\bf base/} containing simulation glue modules with no
  scientific component.
\item {\bf cmod/} containing C code used to accelerate the simulation.
\item {\bf docs/} containing documentation (including auto-generated
  API and user documentation).
\item {\bf gui/} containing graphical user interfaces for simulation
  creation and control.
\item {\bf science/} containing simulation science modules which
  represent a discrete part of an \ao system.
\item {\bf util/} containing utility modules with algorithms required
  by the simulation, which can also be used from a Python terminal by
  the user.
\end{itemize}

\begin{figure}
  \includegraphics[width=0.8\linewidth]{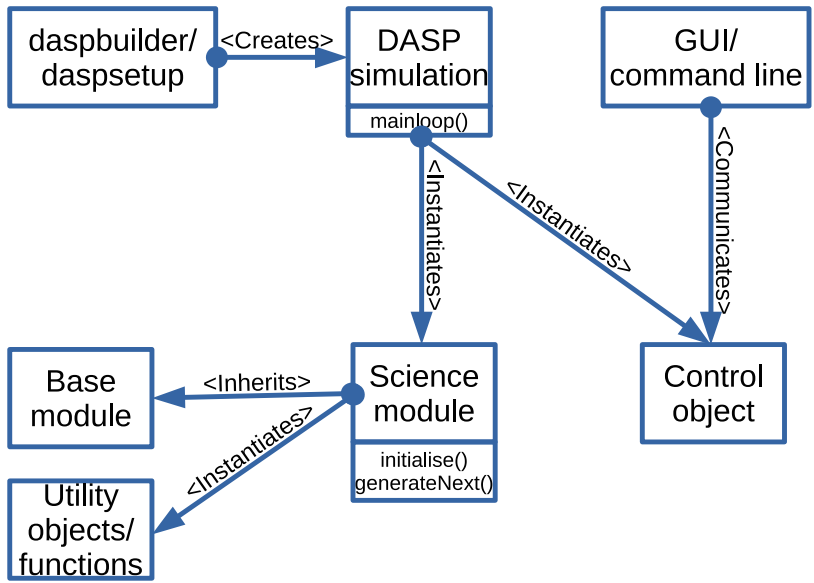}
  \caption{An overview of the DASP architecture, showing simulation
    science modules and their relationship with other DASP
    components.}
  \label{fig:overview}
    
\end{figure}

The science modules are used to form the structure of a simulation,
and represent physical components.  These all inherit from a base
object providing a number of virtual methods to be inherited,
which are called during initialisation and for every iteration of the
running simulation.  The science modules can also import components
from the util/ and cmod/ directories, which provide necessary
algorithms and improved computational performance.  The base directory
also contains glue modules, for example to handle \mpi connection and
modules without scientific content, such as First In First Out (FIFO)
buffers.  Fig.~\ref{fig:overview} provides an overview of the
relationship between \dasp components.

A main \dasp control object is instantiated, parses command line
arguments and loads the specified configuration files.  This object
has a main control loop method, which is passed a list of
instantiated base and science objects.  These are then iterated over,
calling the methods to compute the next iteration for each object.

\subsubsection{Interaction and control}
The control object also includes a communication thread which is used
to allow external (user) connections to the running simulation.  These
connections can then be used to query and modify internal state, for
example to alter the loop gain factor, pause the simulation, or to
extract the current instantaneous \ao system performance.  This
means that \dasp can be used remotely on server computers,
and that multiple simulations can be executed simultaneously, allowing
the user to interact and connect to several simultaneously executing
(independent) simulations.  This allows a parameter space to be
explored more rapidly.  Additionally, multiple users can connect to a
simulation simultaneously, to allow interaction and knowledge sharing.

\subsection{Configuration}
A \dasp simulation structure is stored as an XML file, describing the
links between the different science modules, and relevant
multi-processor information (for example, on which process different
science modules should be run, in the case of a multi-node
simulation).  This XML file is created using daspbuilder (command
line) or daspsetup (graphical), and is then automatically parsed to
generate a Python script.

A \dasp simulation also requires one or more Python-based parameter files, which
specify relevant parameters, dimension the system, and determine the
outputs to be produced.  The parameter file names are passed as command
line parameters.  The use of Python allows dynamic calculation of
variables before execution, e.g.\ automated loading of external data
files, and parsing of contents without requiring additional software tools.

\subsection{Software Functionalities}
\label{sect:func}


The functionality of \dasp is provided by the science modules.  These
include:
\begin{itemize}
\item {\bf Atmospheric phase screen generation:} The atmosphere is
  modelled as a number of thin perturbing layers, which modify
  incident wavefront phase.  A zonal extrusion technique is used
  \citep{Assemat:06} allowing screens of infinite length to be produced.
\item {\bf Pupil phase generation:} The perturbing phase introduced
  along a specified line of sight are identified and summed to give
  the perturbing phase at the telescope pupil, for this particular
  direction.  Within \dasp, both geometric ray tracing and Fresnel
  propagation can be used. Ray tracing is more computationally
  efficient and is the default option, though where high fidelity is
  important (e.g.\ photometry) Fresnel propagation should be selected,
  as scintillation effects are included.
\item {\bf Deformable mirrors:} A deformable mirror is a key component
  in an \ao system, used to mitigate the turbulent phase, with a
  surface shape that is commanded by the \ao control system to match
  the predicted wavefront phase perturbation as closely as possible.
  Both modal and zonal \dm models are available.  Zonal \dm models
  include influence function and interpolated \dms (with spline
  interpolation between actuators to create the \dm surface).  An \ao
  system may contain multiple \dms.  A magic \dm is also available,
  which corrects wavefront phase up to a set modal order without
  requiring a wavefront sensor.
\item {\bf Wavefront sensor and centroider:} A wavefront sensor is
  used to estimate the perturbing wavefront phase along a particular
  line of sight, in the direction of a guide star.  Within \dasp,
  Shack-Hartmann and Pyramid sensors can be used, and include aspects
  such as noise, pixel scale, thresholding and various image
  processing techniques.  The wavefront gradients are estimated by the
  centroiding component, which can use a number of algorithms,
  including centre of gravity or correlation.
\item {\bf Reconstructor:} A wavefront reconstructor module is
  responsible for conversion between wavefront gradient measurements
  to \dm shape.  Multiple reconstructor algorithms are available,
  including dense matrix multiplication (including least squares and
  minimum variance approaches), sparse matrix multiplication using a
  de-densified control matrix (as studied in \citep{basden8}),
  conjugate gradient solvers with preconditioning, and algorithmic
  reconstructors, e.g.\ the Hierarchical Wavefront Reconstructor (HWR)
  \citep{2015MNRAS.448.1199B}.
\item {\bf \Psf generation:} The \psf provides an instantaneous and
  long term (integrated) performance metric along a given line of
  sight, and is essential for diagnosing \ao system capabilities.
\item {\bf Extended object (wide-field) imaging and wavefront
  sensing:} Most \ao simulations consider a single line of sight for
  each wavefront sensor or \psf.  However, \dasp contains an extended
  object module which is able to consider directions within a set
  field of view simultaneously, enabling high fidelity modelling of
  solar \ao systems, and those with extended \lgs spots.
\item {\bf Real-time simulation:} \dasp contains a module which allows
  linkage to a real-time control system, the Durham \ao Real-time
  Controller (DARC) \citep{basden9,basden11}.  This therefore allows on-sky systems
  to be modelled, and algorithms within a real-time control system
  tested and developed, significantly reducing on-sky commissioning
  time.  This hardware-in-the-loop capability is crucial for the
  forthcoming \elts.
\end{itemize}

Additionally, the utility library also provides useful \ao-related
functions, including Zernike mode generation, pixel scale computation,
\psf generation, etc.

\subsection{Parsing results}
Running a large number of simulations will generate a significant
volume of performance data.  Therefore \dasp includes tools to allow
the user to parse these data and extract relevant parameters into
formats that can be used for further processing, analysis and display.


\section{Illustrative Examples}
\label{sect:eg}



\begin{figure}
  \includegraphics[width=0.5\linewidth]{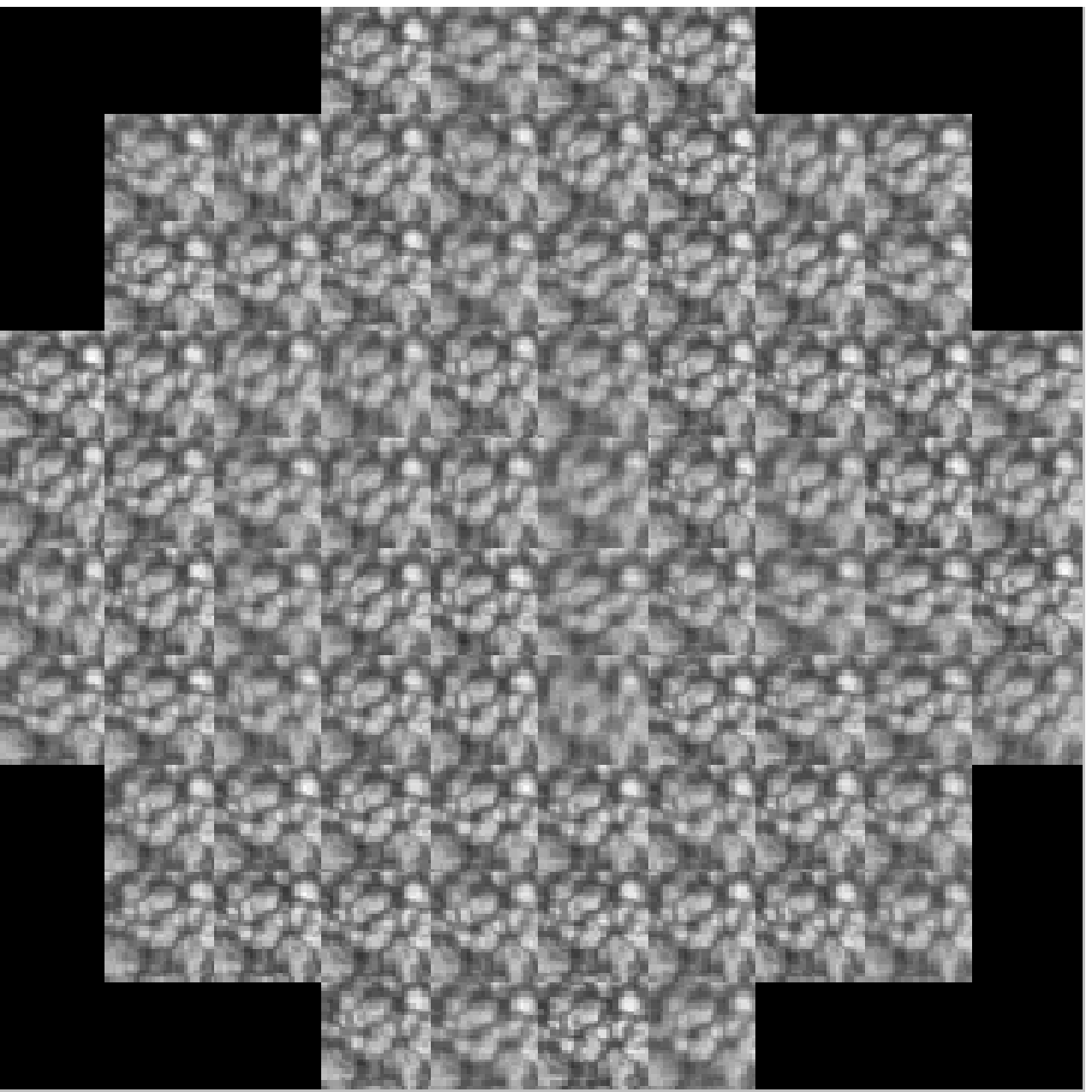}
  \includegraphics[width=0.5\linewidth]{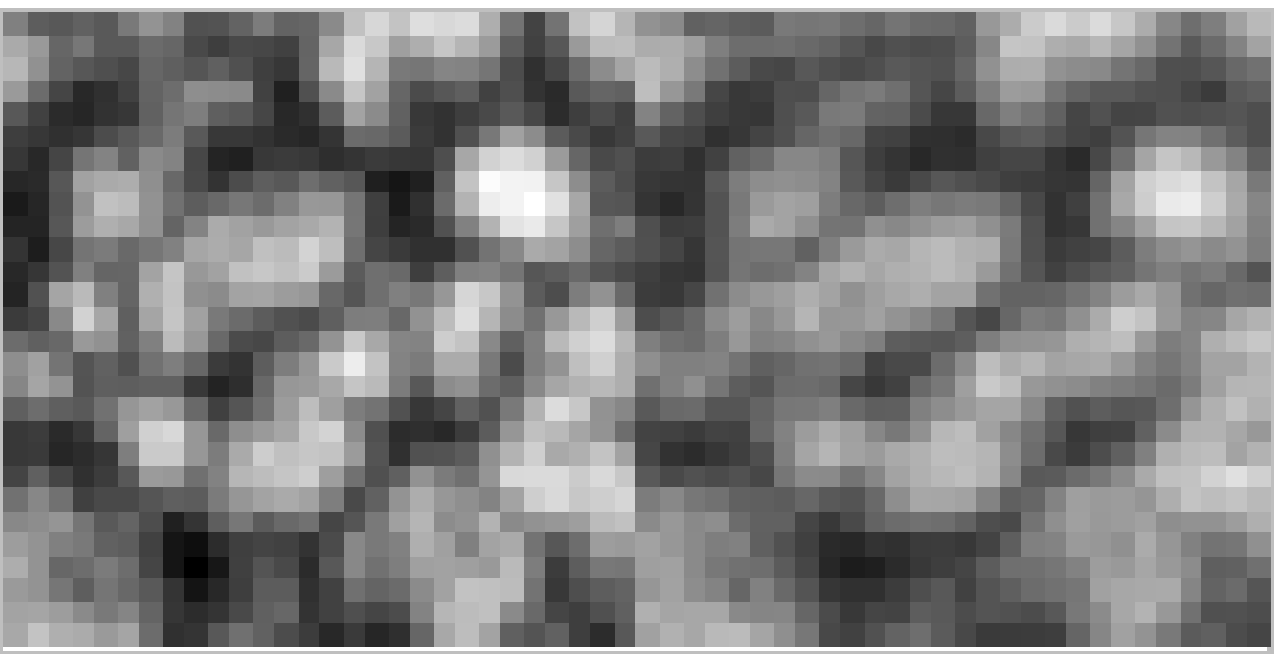}
  \caption{A simulated solar wavefront sensor image showing different
    atmospheric distortion across each sub-aperture.  Two neighbouring
  sub-apertures are also shown in close-up, showing the relative
  distortion between them.}
\label{fig:solar}    
\end{figure}

\dasp provides the ability to flexibly create high fidelity
simulations of \ao systems.  Fig.~\ref{fig:solar} shows a simulated
solar \ao wavefront sensor, demonstrating the extended object
wavefront sensing capability of \dasp.  A video of this wavefront
sensor is also available from SoftwareX.  A closed loop solar \scao
simulation is presented in \ref{sect:solarcode}.  First, the relevant
modules are imported and simulation initialisation is performed,
including instructions to perform initial calibrations (as with a real
\ao system).  The science modules are then instantiated, and finally
the main loop is entered.

This code (along with the associated parameter file) will then
generate an output file which contains \psf information such as the
Strehl ratio, FWHM, ensquared energy, \psf diameter enclosing 50\% of
energy and rms wavefront error.  The \psf itself can also be saved.

\subsection{Hardware-in-the-loop example}
\begin{figure}
  \includegraphics[width=\linewidth]{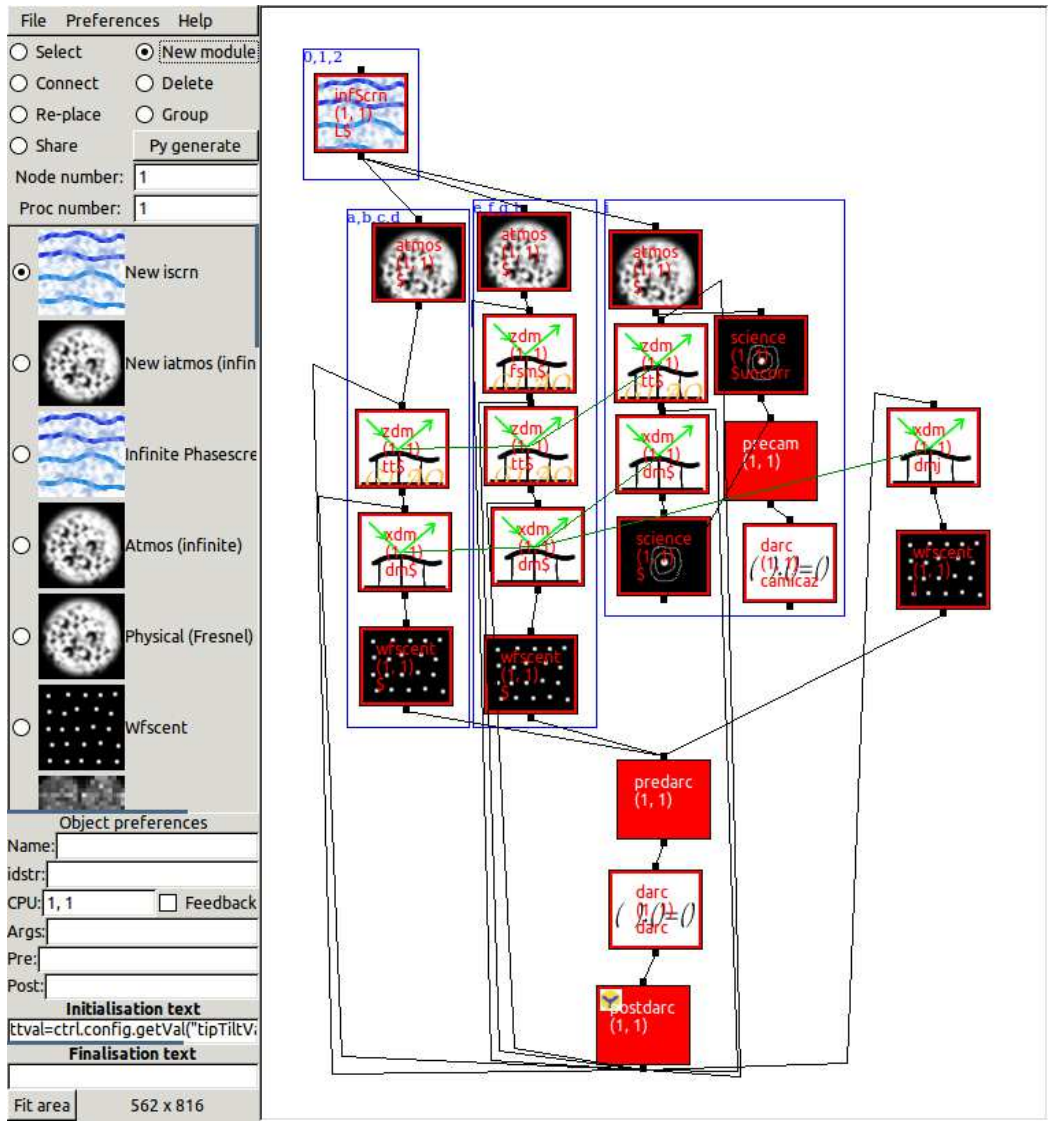}
  \caption{A figure showing the configuration of a
    hardware-in-the-loop simulation used with the  CANARY AO
    instrument to reduce on-sky commissioning time.  This figure is a
    screen-shot of the daspsetup tool and shows how a complex
    real-world simulation can be configured.  The flexibility provided
  by DASP also leaves room for future expansion.  Within this figure,
  one of the modules represents a link to (and from) a real-time
  control system.}
  \label{fig:rtsim}
\end{figure}

As discussed previously, \dasp can be linked with an \ao real-time
control system to provide a hardware-in-the-loop simulation
capability.  Fig.~\ref{fig:rtsim} shows the simulation configuration
that is used with the CANARY \ao instrument \citep{canary}, and is an
illustrative example of how \dasp can be used in real-world situations
to significantly reduce on-sky commissioning time \citep{basden23}.
In this case, \dasp is used to model the atmosphere, telescope, and
\ao system optics, providing input to the CANARY real-time control
system, and updating the model in response to real-time control system
outputs.

\section{Conclusions}
\label{sect:conc}


\dasp is a modular and flexible Monte-Carlo simulation tool for \ao
systems.  It offers high fidelity modelling with good computational
performance.  \dasp is suitable for use at forthcoming extremely large
telescope scales, yet also offers the ability to rapidly prototype and
test new algorithms and concepts.  \dasp has been used for simulation
of both existing and forthcoming \ao systems, and has been verified
against other \ao simulation codes \citep{basden8}.

\section*{Acknowledgements}
This work is funded by the UK Science and Technology Facilities Council
grant ST/K003569/1, consolidated grant ST/L00075X/1 and grant ST/I002871/1.
We thank the referees for their comments which have improved this manuscript.

\appendix

\section{Solar AO system code listing}
\label{sect:solarcode}
A code listing for a solar \scao simulation.
\begin{python}
#Import modules
import science.iscrn
import science.xinterp_dm
import science.wideField
import science.wfscent
import science.tomoRecon
import science.iatmos
import science.science
import util.Ctrl

#Initialisation
ctrl=util.Ctrl.Ctrl(globals=globals())
ctrl.doInitialOpenLoop(startiter=0)
ctrl.initialCommand("wf.control['cal_source']=1",freq=-1,startiter=0)
ctrl.initialCommand("wf.control['cal_source']=0",freq=-1,startiter=1)

#Obtain correlation reference images
ctrl.initialCommand("c.newCorrRef();print 'Done new corr ref'",freq=-1,startiter=1)

#Produce an interaction matrix
ctrl.doInitialPokeThenRun(startiter=2)

#instantiation of science modules:
#Phase screens
iscrn=science.iscrn.iscrn(None,ctrl.config,idstr="allLayers")

#Summed phase along a given direction  
iatmos=science.iatmos.iatmos({"allLayers":iscrn},ctrl.config,idstr="b")

#Wide-field DM (for the wavefront sensor)
dm=science.xinterp_dm.dm(None,ctrl.config,idstr="dma")

#Narrow-field DM (for the PSF generator)
dm2=science.xinterp_dm.dm(None,ctrl.config,idstr="dmNFb")#this one for the science.

#Solar wavefront sensor
wf=science.wideField.WideField({"allLayers":iscrn,"dma":dm},ctrl.config,idstr="a")

#Centroiding object
c=science.wfscent.wfscent(wf,ctrl.config,idstr="acent")

#Wavefront reconstructor
r=science.tomoRecon.recon({"acent":c},ctrl.config,idstr="recon")

#Assign feedback
dm.newParent({"recon":r},"dma")
dm2.newParent({"recon":r,"atmos":iatmos},"dmNFb")

#PSF generator
s=science.science.science(dm2,ctrl.config,idstr="b")

#enter the main loop
execOrder=[iscrn,iatmos,dm,dm2,wf,c,r,s]
ctrl.mainloop(execOrder)
\end{python}

\bibliographystyle{elsarticle-num} 
\bibliography{mybib}







\section*{Current code version}


\begin{table}[!h]
\begin{tabular}{|l|p{6.5cm}|p{6.5cm}|}
\hline
\textbf{Nr.} & \textbf{Code metadata description} & \textbf{Information} \\
\hline
C1 & Current code version &  Commit 3a4faa7 \\
\hline
C2 & Permanent link to code/repository used for this code version & $https://github.com/agb32/dasp$ \\
\hline
C3 & Legal Code License   & GNU Affero General Public Licence v3 \\
\hline
C4 & Code versioning system used & git \\
\hline
C5 & Software code languages, tools, and services used & C, Python, MPI \\
\hline
C6 & Compilation requirements, operating environments \& dependencies
& Linux (full functionality) and OS-X (reduced functionality,
including in some GUI elements)\\
\hline
C7 & If available Link to developer documentation/manual & $http:// dasp.readthedocs.io/en/latest/$ \\
\hline
C8 & Support email for questions & a.g.basden@durham.ac.uk \\
\hline
\end{tabular}
\caption{Code metadata}
\label{hmm} 
\end{table}

\end{document}